\documentstyle[12pt]{article}
\begin{document}
\title{The Mixmaster Universe in Five Dimensions}
\author{Paul Halpern\\Department of Mathematics, Physics and Computer Science\\
University of the Sciences in Philadelphia\\
600 S. 43rd Street\\
Philadelphia, Pa. 19104}
\date{July 15, 2002}
\maketitle
\begin{abstract}
We consider a five dimensional vacuum cosmology with Bianchi type-IX spatial geometry and an extra non-compact coordinate.  Finding a new class of solutions, we examine and rule out the possibility of deterministic chaos.  We interpret this result within the context of induced matter theory.  
\end{abstract}

\section{Introduction}
\label{sec:intro}

The Mixmaster universe, an anisotropic, homogeneous cosmology with Bianchi type-IX spatial geometry, has long been a subject of interest.  Belinskii, Khalatnikov and Lifshitz \cite{bkl1} analyzed its dynamics in their studies of the general approach to the initial singularity.  Misner \cite{misner} proposed a period of Mixmaster dynamics as the possible solution to the horizon problem.

One interesting feature of these models is deterministic chaotic behavior.  This manifests itself in a series of ``Kasner epochs'' in which two of the models' scale factors oscillate, approximated by Kasner solutions linked asymptotically, while the third scale factor monotonically expands.  After a period of time, a so-called ``era,'' the expanding scale factor switches places with one of the oscillating factors -- the second, for instance.  The dynamics then continues with the first and third factors now oscillating, and the second monotonically expanding. This progression from one era to the next occurs indefinitely, with the number of epochs in each successive era appearing as random as a dice toss.  By statistical measures, this represents deterministic chaos, as Barrow \cite{barrow1} has shown and many others have confirmed\cite{szydlowski1,szydlowski2,berger,cornish}.

Chaos is found in the standard 4D vacuum Bianchi type-IX universe, as well as in the vacuum Bianchi type-VIII.  However, for all but a few cases, once matter and energy fields are included (and these are too large to ignore; i.e., not small perturbations), chaotic behavior is suppressed \cite{bkl2,gdl}.  Moreover, as Barrow and Stein-Schabes \cite{barrow2}, as well as Furusawa \cite{furusawa} and others \cite{demianski1,demianski2,biesada} (working independently using various methods) have demonstrated, chaos is absent even in vacuum cases for models of five dimensions or higher.  The absence of higher dimensional chaos includes brane world models, as Coley\cite{coley} has recently determined. 

Might there be a connection between the suppression of chaos in matter-filled 4D models, and the absence of chaos in vacuum 5D (or higher) models?  This linkage can be examined by means of induced matter theory \cite{wesson}, which associates the additional geometrical terms arising from the extra scale factor and derivatives of the fifth coordinate in the five dimensional vacuum Einstein equations with the matter-energy components of four dimensional theory.  In other words, 4D matter stems from 5D geometry.

To investigate this question, we examine the properties of Bianchi type-IX models with the addition of an extended fifth coordinate.  By calculating the density and pressure of the induced matter, we demonstrate that this model cannot be regarded as empty.  Therefore, its properties are qualitatively different from the standard vacuum, non-Kaluza-Klein 4D result, and do not manifest chaos for this reason. 
 
In fact, according to induced matter theory, there is no pure 4D vacuum, as can be seen from the canonical form due to Mashhoon, Liu and Wesson \cite{mashhoon}.  In the 5D Bianchi type-IX case, this precludes chaos, as we demonstrate.

\section{5D Mixmaster Dynamics}
\label{sec:dynamics}

We consider the five dimensional metric:

\begin{equation}
\label{eqn:metric}
ds^2 = e^{\nu} dt^2 - g_{ij} {\omega^i}{\omega^j} - e^{\mu} d{\l}^2 \nonumber
\end{equation}

where the 3D spatial part of the metric can be expressed in diagonal form as:
\begin{equation}
g_{ij} = diag(e^{\alpha},e^{\beta},e^{\gamma})
\end{equation}

The time coordinate $t$ and the three spatial coordinates $x$, $y$ and $z$ have been supplemented with a fifth coordinate $\l$.  We assume that the metric coefficients $\mu$, $\nu$, $\alpha$, $\beta$ and $\gamma$ each depend, in general, on both $t$ and $\l$. 

The one-forms ${\omega^i}$ have the relationship:
\begin{equation}
d{\omega^i} = \frac{1}{2} {C^i}_{jk} {\omega^j}{\omega^k}
\end{equation}
where the non-zero structure constants are ${C^1}_{23}= -{C^1}_{32}= 1$, and cyclic permutations.

The 5D Einstein tensor for the Bianchi type-IX case can be written in the form:
\begin{eqnarray}
\label{eqn:ein1}
G^0_0= e^{-\nu} ( & & -\frac{1}{4}{{\mu}^.}{{\alpha}^.}-\frac{1}{4}{{\mu}^.}{{\beta}^.}-\frac{1}{4}{{\mu}^.}{{\gamma}^.}-\frac{1}{4}{{\alpha}^.}{{\beta}^.}-\frac{1}{4}{{\alpha}^.}{{\gamma}^.}-\frac{1}{4}{{\beta}^.}{{\gamma}^.})+ \nonumber \\ e^{-\mu} ( & &\frac{1}{2} {\alpha}^{**} +\frac{1}{2} {\beta}^{**}+ \frac{1}{2} {\gamma}^{**} + \frac{1}{4} {{\alpha}^*}^2 + \frac{1}{4} {{\beta}^*}^2 + \frac{1}{4} {{\gamma}^*}^2- \nonumber \\& & \frac{1}{4}{{\mu}^*}{{\alpha}^*} -\frac{1}{4}{{\mu}^*}{{\beta}^*}-\frac{1}{4}{{\mu}^*}{{\gamma}^*}+\frac{1}{4}{{\alpha}^*}{{\beta}^*}+\frac{1}{4}{{\alpha}^*}{{\gamma}^*}+\frac{1}{4}{{\beta}^*}{{\gamma}^*}) + \nonumber \\ & &\frac{e^{\alpha}}{4 e^{\beta} e^{\gamma}} + \frac{e^{\beta}}{4 e^{\alpha} e^{\gamma}} + \frac{e^{\gamma}}{4 e^{\alpha} e^{\beta}} - \frac{1}{2 e^{\alpha}} - \frac{1}{2 e^{\beta}} - \frac{1}{2 e^{\gamma}} \\ \nonumber \\ 
\label{eqn:ein2}
G^0_4= e^{-\nu} ( & & 2{\alpha^{.*}} + 2{\beta^{.*}} + 2{\gamma^{.*}} +{{\alpha}^.}{{\alpha}^*} + {{\beta}^.}{{\beta}^*} + {{\gamma}^.}{{\gamma}^*} - \nonumber \\ & & {{\alpha}^.}{{\nu}^*}- {{\beta}^.}{{\nu}^*}- {{\gamma}^.}{{\nu}^*} - {{\alpha}^*}{{\mu}^.}- {{\beta}^*}{{\mu}^.}- {{\gamma}^*}{{\mu}^.})\\ \nonumber \\ 
\label{eqn:ein3}
G^1_1=   e^{-\nu} ( & & -\frac{1}{2}{{\beta}^{..}}-\frac{1}{2}{{\gamma}^{..}}-\frac{1}{2}{{\mu}^{..}}-\frac{1}{4}{{\beta}^.}^2-\frac{1}{4}{{\gamma}^.}^2-\frac{1}{4}{{\mu}^.}^2- \nonumber \\ & & \frac{1}{4}{{\beta}^.}{{\gamma}^.} -\frac{1}{4}{{\beta}^.}{{\mu}^.} - \frac{1}{4}{{\gamma}^.}{{\mu}^.} + \frac{1}{4}{{\beta}^.}{{\nu}^.} +\frac{1}{4}{{\gamma}^.}{{\nu}^.} + \frac{1}{4}{{\mu}^.}{{\nu}^.}) + \nonumber \\ e^{-\mu} ( & & \frac{1}{2}{{\beta}^{**}} + \frac{1}{2}{{\gamma}^{**}} + \frac{1}{2}{{\nu}^{**}} + \frac{1}{4}{{\beta}^*}^2 +\frac{1}{4}{{\gamma}^*}^2 + \frac{1}{4}{{\nu}^*}^2+ \nonumber \\ & & \frac{1}{4} {{\beta}^*}{{\gamma}^*} + \frac{1}{4} {{\beta}^*}{{\nu}^*} +\frac{1}{4} {{\gamma}^*}{{\nu}^*} - \frac{1}{4}{{\beta}^*}{{\mu}^*} - \frac{1}{4}{{\gamma}^*}{{\mu}^*} - \frac{1}{4}{{\mu}^*}{{\nu}^*}) + \nonumber \\ & &\frac{3 e^{\alpha}}{4 e^{\beta} e^{\gamma}} - \frac{e^{\beta}}{4 e^{\alpha} e^{\gamma}} - \frac{e^{\gamma}}{4 e^{\alpha} e^{\beta}} + \frac{1}{2 e^{\alpha}} - \frac{1}{2 e^{\beta}} - \frac{1}{2 e^{\gamma}} \\ \nonumber \\
\label{eqn:ein4}
G^2_2=   e^{-\nu} ( & & -\frac{1}{2}{{\alpha}^{..}}-\frac{1}{2}{{\gamma}^{..}}-\frac{1}{2}{{\mu}^{..}}-\frac{1}{4}{{\alpha}^.}^2-\frac{1}{4}{{\gamma}^.}^2-\frac{1}{4}{{\mu}^.}^2- \nonumber \\ & & \frac{1}{4}{{\alpha}^.}{{\gamma}^.} -\frac{1}{4}{{\alpha}^.}{{\mu}^.} - \frac{1}{4}{{\gamma}^.}{{\mu}^.} + \frac{1}{4}{{\alpha}^.}{{\nu}^.} +\frac{1}{4}{{\gamma}^.}{{\nu}^.} + \frac{1}{4}{{\mu}^.}{{\nu}^.}) + \nonumber \\ e^{-\mu} ( & & \frac{1}{2}{{\alpha}^{**}} + \frac{1}{2}{{\gamma}^{**}} + \frac{1}{2}{{\nu}^{**}} + \frac{1}{4}{{\alpha}^*}^2 +\frac{1}{4}{{\gamma}^*}^2 + \frac{1}{4}{{\nu}^*}^2+ \nonumber \\ & & \frac{1}{4} {{\alpha}^*}{{\gamma}^*} + \frac{1}{4} {{\alpha}^*}{{\nu}^*} +\frac{1}{4} {{\gamma}^*}{{\nu}^*} - \frac{1}{4}{{\alpha}^*}{{\mu}^*} - \frac{1}{4}{{\gamma}^*}{{\mu}^*} - \frac{1}{4}{{\mu}^*}{{\nu}^*}) - \nonumber \\ & &\frac{e^{\alpha}}{4 e^{\beta} e^{\gamma}} + \frac{3 e^{\beta}}{4 e^{\alpha} e^{\gamma}} - \frac{e^{\gamma}}{4 e^{\alpha} e^{\beta}} - \frac{1}{2 e^{\alpha}} + \frac{1}{2 e^{\beta}} - \frac{1}{2 e^{\gamma}}\\ \nonumber \\
\label{eqn:ein5}
G^3_3=   e^{-\nu} ( & & -\frac{1}{2}{{\alpha}^{..}}-\frac{1}{2}{{\beta}^{..}}-\frac{1}{2}{{\mu}^{..}}-\frac{1}{4}{{\alpha}^.}^2-\frac{1}{4}{{\beta}^.}^2-\frac{1}{4}{{\mu}^.}^2- \nonumber \\ & & \frac{1}{4}{{\alpha}^.}{{\beta}^.} -\frac{1}{4}{{\alpha}^.}{{\mu}^.} - \frac{1}{4}{{\beta}^.}{{\mu}^.} + \frac{1}{4}{{\alpha}^.}{{\nu}^.} +\frac{1}{4}{{\beta}^.}{{\nu}^.} + \frac{1}{4}{{\mu}^.}{{\nu}^.}) + \nonumber \\ e^{-\mu} ( & & \frac{1}{2}{{\alpha}^{**}} + \frac{1}{2}{{\beta}^{**}} + \frac{1}{2}{{\nu}^{**}} + \frac{1}{4}{{\alpha}^*}^2 +\frac{1}{4}{{\beta}^*}^2 + \frac{1}{4}{{\nu}^*}^2+ \nonumber \\ & & \frac{1}{4} {{\alpha}^*}{{\beta}^*} + \frac{1}{4} {{\alpha}^*}{{\nu}^*} +\frac{1}{4} {{\beta}^*}{{\nu}^*} - \frac{1}{4}{{\alpha}^*}{{\mu}^*} - \frac{1}{4}{{\beta}^*}{{\mu}^*} - \frac{1}{4}{{\mu}^*}{{\nu}^*}) - \nonumber \\ & &\frac{e^{\alpha}}{4 e^{\beta} e^{\gamma}} - \frac{e^{\beta}}{4 e^{\alpha} e^{\gamma}} + \frac{3 e^{\gamma}}{4 e^{\alpha} e^{\beta}} - \frac{1}{2 e^{\alpha}} - \frac{1}{2 e^{\beta}} + \frac{1}{2 e^{\gamma}}\\ \nonumber \\
\label{eqn:ein6}
G^4_4=   e^{-\nu} ( & & -\frac{1}{2}{{\alpha}^{..}} -\frac{1}{2}{{\beta}^{..}}-\frac{1}{2}{{\gamma}^{..}} -\frac{1}{4}{{\alpha}^.}^2-\frac{1}{4}{{\beta}^.}^2-\frac{1}{4}{{\gamma}^.}^2- \nonumber \\ & & \frac{1}{4}{{\alpha}^.}{{\beta}^.} -\frac{1}{4}{{\alpha}^.}{{\gamma}^.} - \frac{1}{4}{{\beta}^.}{{\gamma}^.} + \frac{1}{4}{{\alpha}^.}{{\nu}^.} +\frac{1}{4}{{\beta}^.}{{\nu}^.} + \frac{1}{4}{{\gamma}^.}{{\nu}^.}) + \nonumber \\ e^{-\mu} ( & & \frac{1}{4} {{\alpha}^*}{{\beta}^*} + \frac{1}{4} {{\alpha}^*}{{\gamma}^*} +\frac{1}{4} {{\beta}^*}{{\gamma}^*} +\frac{1}{4} {{\alpha}^*}{{\nu}^*} + \frac{1}{4} {{\beta}^*}{{\nu}^*} + \frac{1}{4}{{\gamma}^*}{{\nu}^*})+ \nonumber \\  & &\frac{e^{\alpha}}{4 e^{\beta} e^{\gamma}} + \frac{e^{\beta}}{4 e^{\alpha} e^{\gamma}} + \frac{e^{\gamma}}{4 e^{\alpha} e^{\beta}} - \frac{1}{2 e^{\alpha}} - \frac{1}{2 e^{\beta}} - \frac{1}{2 e^{\gamma}}
\end{eqnarray}

where we use overdots to represent partial derivatives with respect to $t$, and asterisks to represent partial derivatives with respect to $\l$.

Following the standard procedure, we collect each of the terms in the 5D vacuum Einstein equations, ${G^{\mu}}_{\nu} = 0$,  dependent on either $\mu$ or on derivatives with respect to $l$, place them on the right-hand side and identify them as the induced matter density and pressure components:

\begin{eqnarray}
\label{eqn:density}
8 \pi \rho= e^{-\nu} ( & & -\frac{1}{4} {{\alpha}^.}{{\mu}^.}- \frac{1}{4} {{\beta}^.}{{\mu}^.}- \frac{1}{4} {{\gamma}^.}{{\mu}^.}) + \nonumber \\ e^{-\mu} ( & & \frac{1}{2} {\alpha}^{**} +\frac{1}{2} {\beta}^{**}+ \frac{1}{2} {\gamma}^{**} + \frac{1}{4} {{\alpha}^*}^2 + \frac{1}{4} {{\beta}^*}^2 + \frac{1}{4} {{\gamma}^*}^2- \nonumber \\ & & \frac{1}{4}{{\mu}^*}{{\alpha}^*}-\frac{1}{4}{{\mu}^*}{{\beta}^*}-\frac{1}{4}{{\mu}^*}{{\gamma}^*}+\frac{1}{4}{{\alpha}^*}{{\beta}^*}+\frac{1}{4}{{\alpha}^*}{{\gamma}^*}+\frac{1}{4}{{\beta}^*}{{\gamma}^*})\\ \nonumber \\
\label{eqn:press1}
8 \pi P_1 = e^{-\nu} ( & & \frac{1}{2}{{\mu}^{..}}+ \frac{1}{4}{{\mu}^.}^2+ \frac{1}{4} {{\beta}^.}{{\mu}^.}+ \frac{1}{4} {{\gamma}^.}{{\mu}^.}- \frac{1}{4} {{\mu}^.}{{\nu}^.}) + \nonumber \\ e^{-\mu} ( & & -\frac{1}{2}{{\beta}^{**}} - \frac{1}{2}{{\gamma}^{**}} -\frac{1}{2}{{\nu}^{**}} - \frac{1}{4}{{\beta}^*}^2 -\frac{1}{4}{{\gamma}^*}^2 - \frac{1}{4}{{\nu}^*}^2- \nonumber \\ & & \frac{1}{4} {{\beta}^*}{{\gamma}^*} - \frac{1}{4} {{\beta}^*}{{\nu}^*} -\frac{1}{4} {{\gamma}^*}{{\nu}^*} + \frac{1}{4}{{\beta}^*}{{\mu}^*} + \frac{1}{4}{{\gamma}^*}{{\mu}^*} + \frac{1}{4}{{\mu}^*}{{\nu}^*})\\ \nonumber \\
\label{eqn:press2}
8 \pi P_2 = e^{-\nu} ( & & \frac{1}{2}{{\mu}^{..}}+ \frac{1}{4}{{\mu}^.}^2+ \frac{1}{4} {{\alpha}^.}{{\mu}^.}+ \frac{1}{4} {{\gamma}^.}{{\mu}^.}- \frac{1}{4} {{\mu}^.}{{\nu}^.}) +\nonumber \\ e^{-\mu} ( & & -\frac{1}{2}{{\alpha}^{**}} - \frac{1}{2}{{\gamma}^{**}} - \frac{1}{2}{{\nu}^{**}} - \frac{1}{4}{{\alpha}^*}^2 -\frac{1}{4}{{\gamma}^*}^2 - \frac{1}{4}{{\nu}^*}^2- \nonumber \\ & & \frac{1}{4} {{\alpha}^*}{{\gamma}^*} - \frac{1}{4} {{\alpha}^*}{{\nu}^*} -\frac{1}{4} {{\gamma}^*}{{\nu}^*} + \frac{1}{4}{{\alpha}^*}{{\mu}^*} + \frac{1}{4}{{\gamma}^*}{{\mu}^*} + \frac{1}{4}{{\mu}^*}{{\nu}^*})\\ \nonumber \\
\label{eqn:press3}
8 \pi P_3 = e^{-\nu} ( & & \frac{1}{2}{{\mu}^{..}}+ \frac{1}{4}{{\mu}^.}^2+ \frac{1}{4} {{\alpha}^.}{{\mu}^.}+ \frac{1}{4} {{\beta}^.}{{\mu}^.}- \frac{1}{4} {{\mu}^.}{{\nu}^.}) + \nonumber \\ e^{-\mu} ( & & -\frac{1}{2}{{\alpha}^{**}} - \frac{1}{2}{{\beta}^{**}} - \frac{1}{2}{{\nu}^{**}} - \frac{1}{4}{{\alpha}^*}^2 -\frac{1}{4}{{\beta}^*}^2 - \frac{1}{4}{{\nu}^*}^2- \nonumber \\ & & \frac{1}{4} {{\alpha}^*}{{\beta}^*} - \frac{1}{4} {{\alpha}^*}{{\nu}^*} -\frac{1}{4} {{\beta}^*}{{\nu}^*} + \frac{1}{4}{{\alpha}^*}{{\mu}^*} + \frac{1}{4}{{\beta}^*}{{\mu}^*} + \frac{1}{4}{{\mu}^*}{{\nu}^*})
\end{eqnarray}

The Einstein equations now can be written in 4D form as:

\begin{eqnarray}
\label{eqn:einstein0}
e^{-\nu} ( & & -\frac{1}{4}{{\alpha}^.}{{\beta}^.}-\frac{1}{4}{{\alpha}^.}{{\gamma}^.}-\frac{1}{4}{{\beta}^.}{{\gamma}^.})+ \nonumber \\  & &\frac{e^{\alpha}}{4 e^{\beta} e^{\gamma}} + \frac{e^{\beta}}{4 e^{\alpha} e^{\gamma}} + \frac{e^{\gamma}}{4 e^{\alpha} e^{\beta}} - \frac{1}{2 e^{\alpha}} - \frac{1}{2 e^{\beta}} - \frac{1}{2 e^{\gamma}}+ 8 \pi \rho = 0\\ \nonumber \\ 
\label{eqn:einstein1}
e^{-\nu} ( & & -\frac{1}{2}{{\beta}^{..}}-\frac{1}{2}{{\gamma}^{..}}-\frac{1}{4}{{\beta}^.}^2-\frac{1}{4}{{\gamma}^.}^2- \frac{1}{4}{{\beta}^.}{{\gamma}^.} + \frac{1}{4}{{\beta}^.}{{\nu}^.} +\frac{1}{4}{{\gamma}^.}{{\nu}^.})+ \nonumber \\ & &\frac{3 e^{\alpha}}{4 e^{\beta} e^{\gamma}} - \frac{e^{\beta}}{4 e^{\alpha} e^{\gamma}} - \frac{e^{\gamma}}{4 e^{\alpha} e^{\beta}} + \frac{1}{2 e^{\alpha}} - \frac{1}{2 e^{\beta}} - \frac{1}{2 e^{\gamma}} - 8 \pi P_1 = 0 \\ \nonumber \\
\label{eqn:einstein2}
e^{-\nu} ( & & -\frac{1}{2}{{\alpha}^{..}}-\frac{1}{2}{{\gamma}^{..}}- \frac{1}{4}{{\alpha}^.}^2-\frac{1}{4}{{\gamma}^.}^2- \frac{1}{4}{{\alpha}^.}{{\gamma}^.}  + \frac{1}{4}{{\alpha}^.}{{\nu}^.} +\frac{1}{4}{{\gamma}^.}{{\nu}^.}) - \nonumber \\ & &\frac{e^{\alpha}}{4 e^{\beta} e^{\gamma}} + \frac{3 e^{\beta}}{4 e^{\alpha} e^{\gamma}} - \frac{e^{\gamma}}{4 e^{\alpha} e^{\beta}} - \frac{1}{2 e^{\alpha}} + \frac{1}{2 e^{\beta}} - \frac{1}{2 e^{\gamma}} - 8 \pi P_2 = 0 \\ \nonumber \\ 
\label{eqn:einstein3}
e^{-\nu} ( & & -\frac{1}{2}{{\alpha}^{..}}-\frac{1}{2}{{\beta}^{..}}-\frac{1}{4}{{\alpha}^.}^2-\frac{1}{4}{{\beta}^.}^2- \frac{1}{4}{{\alpha}^.}{{\beta}^.} + \frac{1}{4}{{\alpha}^.}{{\nu}^.} +\frac{1}{4}{{\beta}^.}{{\nu}^.}) - \nonumber \\ & &\frac{e^{\alpha}}{4 e^{\beta} e^{\gamma}} - \frac{e^{\beta}}{4 e^{\alpha} e^{\gamma}} + \frac{3 e^{\gamma}}{4 e^{\alpha} e^{\beta}} - \frac{1}{2 e^{\alpha}} - \frac{1}{2 e^{\beta}} + \frac{1}{2 e^{\gamma}} - 8 \pi P_3 = 0 \\ \nonumber \\
\label{eqn:einstein4}
e^{-\nu} ( & & 2{\alpha^{.*}} + 2{\beta^{.*}} + 2{\gamma^{.*}} +{{\alpha}^.}{{\alpha}^*} + {{\beta}^.}{{\beta}^*} + {{\gamma}^.}{{\gamma}^*} - \nonumber \\ & & {{\alpha}^.}{{\nu}^*}- {{\beta}^.}{{\nu}^*}- {{\gamma}^.}{{\nu}^*} - {{\alpha}^*}{{\mu}^.}- {{\beta}^*}{{\mu}^.}- {{\gamma}^*}{{\mu}^.}) = 0\\ \nonumber \\ 
\label{eqn:einstein5}
e^{-\nu} ( & & -\frac{1}{2}{{\alpha}^{..}} -\frac{1}{2}{{\beta}^{..}}-\frac{1}{2}{{\gamma}^{..}} -\frac{1}{4}{{\alpha}^.}^2-\frac{1}{4}{{\beta}^.}^2-\frac{1}{4}{{\gamma}^.}^2- \nonumber \\ & & \frac{1}{4}{{\alpha}^.}{{\beta}^.} -\frac{1}{4}{{\alpha}^.}{{\gamma}^.} - \frac{1}{4}{{\beta}^.}{{\gamma}^.} + \frac{1}{4}{{\alpha}^.}{{\nu}^.} +\frac{1}{4}{{\beta}^.}{{\nu}^.} + \frac{1}{4}{{\gamma}^.}{{\nu}^.}) + \nonumber \\ e^{-\mu} ( & & \frac{1}{4} {{\alpha}^*}{{\beta}^*} + \frac{1}{4} {{\alpha}^*}{{\gamma}^*} +\frac{1}{4} {{\beta}^*}{{\gamma}^*} +\frac{1}{4} {{\alpha}^*}{{\nu}^*} + \frac{1}{4} {{\beta}^*}{{\nu}^*} + \frac{1}{4}{{\gamma}^*}{{\nu}^*}) + \nonumber \\  & &\frac{e^{\alpha}}{4 e^{\beta} e^{\gamma}} + \frac{e^{\beta}}{4 e^{\alpha} e^{\gamma}} + \frac{e^{\gamma}}{4 e^{\alpha} e^{\beta}} - \frac{1}{2 e^{\alpha}} - \frac{1}{2 e^{\beta}} - \frac{1}{2 e^{\gamma}} = 0
\end{eqnarray} 

We solve equations (\ref{eqn:einstein0}-\ref{eqn:einstein5}) through the method of separation of variables.  We write:

\begin{eqnarray}
e^{\alpha} &=& [a(t){\hat a(l)}]^2 \\
e^{\beta} &=& [b(t){\hat b(l)}]^2  \\
e^{\gamma} &=& [c(t){\hat c(l)}]^2 \\
e^{\mu} &=& [d(t){\hat d(l)}]^2 \\
e^{\nu} &=& [f(t) {\hat f(l)}]^2
\end{eqnarray} 
where we have explicitly indicated dependence on either $t$ or $l$ exclusively.

We then make the following simplifying assumptions:
\begin{eqnarray}
f &=& a b c d\\
{\hat a}&=&e^{s_1 l}  \\
{\hat b}&=&e^{s_2 l}  \\
{\hat c}&=&e^{s_3 l}  \\
{\hat d}&=&e^{s_4 l}  \\
{\hat f}&=&e^{s_0 l}  
\end{eqnarray}
where the $s_i$ are all constants.

To solve, we impose the conditions:
\begin{eqnarray}
\label{eqn:s1}
s_1 &=& 0\\
\label{eqn:s0}
s_0 &=& s_2 + s_3
\end{eqnarray}

Equations (\ref{eqn:einstein1}-\ref{eqn:einstein3}) then reduce to:
\begin{eqnarray}
\label{eqn:a}
{(\frac{a^{.}}{a})}^{.} &=& \frac{1}{2} ((b^2 e^{2s_2 l} - c^2 e^{2s_3 l})^2 - a^4) d^2 \\
\label{eqn:b}
{(\frac{b^{.}}{b})}^{.} &=& \frac{1}{2} ((c^2 e^{2s_3 l} - a^2)^2 -  b^4 e^{4 s_2}) d^2 + \nonumber \\ & & (2{s_2}^2+ 2 s_2 s_3 - s_2 s_4) a^2 b^2 c^2 e^{2(s2+s3-s4)l}\\
\label{eqn:c}
{(\frac{c^{.}}{c})}^{.} &=& \frac{1}{2} ((a^2 - b^2 e^{2s_2})^2 - c^4 e^{4 s_3}) d^2 + \nonumber \\ & & (2{s_3}^2+ 2 s_2 s_3 - s_3 s_4) a^2 b^2 c^2 e^{2(s2+s3-s4)l}
\end{eqnarray}

Following the procedure of Belinskii, Khalatnikov and Lifshitz, we examine this system's evolution backward in time. If one could neglect all the terms on the RHS of equations (\ref{eqn:a}-\ref{eqn:c}) then one would obtain a 5D generalized Kasner solution, similar to that found by one of us in an earlier paper \cite{halpern}, with:
\begin{eqnarray}
\label{eqn:kas1}
a &\sim &e^{\Lambda p_1 t} \\
\label{eqn:kas2}
b &\sim &e^{\Lambda p_2 t} \\
\label{eqn:kas3}
c &\sim &e^{\Lambda p_3 t}
\end{eqnarray}
where $\Lambda$ and $p_i$ (the generalized Kasner exponents) are constants.

The relationship amongst the generalized Kasner exponents is:
\begin{eqnarray}
\label{eqn:p1}
&&\sum_{i=1}^{4}{p_i}=1\\
\label{eqn:p2}
&&\sum_{i=1}^{4}{p_i}^2=1
\end{eqnarray} 

Suppose, as these constraints allow, one of the three spatial exponents is negative--$p_1$, for instance--and the other two, $p_2$ and $p_3$, are positive.  In that case $a$ would increase as the system evolves backward in time, and $b$ and $c$ would decrease.  Soon the RHS of equations (\ref{eqn:a}-\ref{eqn:c}) would become dominated by the terms containing $a^4$, and they would reduce to the approximate form:
\begin{eqnarray}
\label{eqn:a-reduced}
{(\frac{a^{.}}{a})}^{.} &=& - \frac{1}{2} a^4 d^2 \\
\label{eqn:b-reduced}
{(\frac{b^{.}}{b})}^{.} &=& \frac{1}{2} a^4 d^2 \\
\label{eqn:c-reduced}
{(\frac{c^{.}}{c})}^{.} &=& \frac{1}{2} a^4 d^2
\end{eqnarray}

In addition, equation (\ref{eqn:einstein5}) can be written as:
\begin{equation}
\label{eqn:d}
{(\frac{d^{.}}{d})}^{.} = 2 ({s_2}^2+{s_3}^2+ {s_2}{s_3}- {s_2}{s_4}-{s_3}{s_4}) e^{2(s_4-s_2-s_3)l} a^2 b^2 c^2
\end{equation}

which yields:
\begin{eqnarray}
\label{eqn:d-solved}
d &=&e^{\Lambda p_4 t} \\
\label{eqn:s4}
s_4 &=& \frac{{s_2}^2+ {s_2}{s_3}+ {s_3}^2}{s_2+ s_3}
\end{eqnarray}

where $\Lambda$ is a constant.

Substituting (\ref{eqn:d-solved}) into (\ref{eqn:a-reduced}-\ref{eqn:c-reduced}), and then solving along with equation (\ref{eqn:einstein0}) yields:
\begin{eqnarray}
a &=& \sqrt {{\frac {\Lambda\, \left( 2\,{p_1}+{p_4} \right) }{\cosh
 \left( \Lambda\, \left( 2\,{p_1}+{p_4} \right) t \right) }}}{e^
{-1/2\,\Lambda\,{p_4}\,t}}\\
b &=& {b_0}\,\sqrt {\cosh \left( \Lambda\, \left( 2\,{p_1}+{p_4}
 \right) t \right) }{e^{-1/2\,\Lambda\,t \left( -2\,{p_2}+2\,{p_1}+{p_4} \right) }}\\
c &=& {c_0}\,\sqrt {\cosh \left( \Lambda\, \left( 2\,{p_1}+{p_4}
 \right) t \right) }{e^{-1/2\,\Lambda\,t \left( -2\,{p_3}+2\,{p_1}+{p_4} \right) }}
\end{eqnarray}

where $b_0$ and $c_0$ are constants.  The choice of remaining constants is such that for large $t$ the solution asymptotically matches the generalized Kasner behavior described by (\ref{eqn:kas1}-\ref{eqn:kas3}).

Taken into account conditions (\ref{eqn:s1}, \ref{eqn:s0} and \ref{eqn:s4}), as well as equation (\ref{eqn:einstein4}), we can now express the full solution as:
\begin{eqnarray}
e^{\nu} &=& \Lambda (2p_1+p_4){b_0}^2{c_0}^2 \cosh{(\Lambda (2p_1+p_4)t)} \nonumber \\ & & e^{(\Lambda(-4p_1+2p_2+2p_3-p_4)t + 2s_3(1-\kappa)l)}\\
\label{eqn:alpha}
e^{\alpha} &=& {\frac {\Lambda\, \left( 2\,{p_1}+{p_4} \right) {e^{-\Lambda\,{
p_4}\,t}}}{\cosh \left( \Lambda\, \left( 2\,{p_1}+{p_4}
 \right) t \right) }}\\
\label{eqn:beta}
e^{\beta} &=& {{b_0}}^{2}\cosh \left( \Lambda\, \left( 2\,{p_1}+{p_4}
 \right) t \right) {e^{(2\,\Lambda\,t{p_2}-2\,\Lambda\,t{p_1}-
\Lambda\,{p_4}\,t-2\, \kappa {s_3}\,l)}}\\
\label{eqn:gamma}
e^{\gamma} &=& {{c_0}}^{2}\cosh \left( \Lambda\, \left( 2\,{p_1}+{p_4}
 \right) t \right) {e^{(2\,\Lambda\,t{p_3}-2\,\Lambda\,t{p_1}-
\Lambda\,{p_4}\,t+2\,{s_3}\,l)}}\\
e^{\mu} &=& {e^{(2\,\Lambda\,{p_4}\,t+4\,{s_3}\,(1-\kappa)\,l\,
)}}
\end{eqnarray}

where $\kappa = \frac{3}{2} \pm \frac{1}{2}\sqrt{5}$.
 
Note that for $s_3=0$, dependency on $l$ vanishes, and this solution reduces to that of Barrow and Stein-Schabes \cite{barrow2}.  Otherwise, for non-vanishing $s_3$, we obtain the additional constraint:
\begin{equation}
\label{eqn:p3}
p_1 - p_1 \kappa - p_2 + p_3 \kappa = 0
\end{equation} 

Let's now examine the epoch to epoch behavior of this model.  For $t \rightarrow \infty $ its spatial scale factors asymptotically match the generalized Kasner solution:

\begin{eqnarray}
\alpha &=& 2 \Lambda p_1 t\\
\beta &=& 2 \Lambda p_2 t - 2 \kappa s_3 l \\
\gamma &=& 2 \Lambda p_3 t + 2 s_3 l \\
\end{eqnarray} 

Then, as $t$ runs backward, its scale factors evolve according to (\ref{eqn:alpha} - \ref{eqn:gamma}).  Taking $t \rightarrow -\infty $ ($t$ in this coordinate system ranges from -$\infty$ to $\infty$), its spatial scale factors asymptotically match another generalized Kasner solution:

\begin{eqnarray}
{\alpha}^{\prime} &=& 2 {\Lambda}^{\prime} {p_1}^{\prime} t\\
{\beta}^{\prime} &=& 2 {\Lambda}^{\prime} {p_2}^{\prime} t - 2 \kappa s_3 l \\
{\gamma}^{\prime} &=& 2 {\Lambda}^{\prime} {p_3}^{\prime} t + 2 s_3 l \\
\end{eqnarray} 

where the old and new epochs are connected by the relationships:

\begin{eqnarray} 
{\Lambda}^{\prime} {p_1}^{\prime} &=& - \Lambda (p_1 + p_4) \\
{\Lambda}^{\prime} {p_2}^{\prime} &=& \Lambda (p_2 - 2 p_1 - p_4)\\
{\Lambda}^{\prime} {p_3}^{\prime} &=& \Lambda (p_3 - 2 p_1 - p_4)\\
{\Lambda}^{\prime} {p_4}^{\prime} &=& \Lambda p_4
\end{eqnarray}

These epoch-epoch transitions continue until $p_1$,$p_2$ and $p_3$ are all positive. Such a circumstance is permitted by conditions (\ref{eqn:p1}, \ref{eqn:p2} and \ref{eqn:p3}) for a finite range of generalised Kasner exponents. When that happens, the RHS of equations (\ref{eqn:a}-\ref{eqn:c}) vanish, and no more transitions take place.  The universe continues to evolve indefinitely with the same set of exponents.  Because the number of transitions is finite, this solution is manifestly not chaotic.

\section{Induced Matter Properties}
\label{sec:induced}

Let's now examine the density and pressure of the induced matter associated with this solution.  Substituting into eqns. (\ref{eqn:density}-\ref{eqn:press3}), we obtain:

\begin{eqnarray}
8 \pi \rho &=& (3p_4+\sqrt {-20\,{{p_4}}^{2}+10\,{p_4}+10})g_1 - \nonumber \\ & & (\frac{p_4}{3}+\frac{2}{3}+\frac{\sqrt {-20\,{{p_4}}^{2}+10\,{p_4}+10}}{3})g_2\\
8 \pi P_1 &=& p_4 g_1 + (\frac{p_4}{3}+\frac{2}{3}+\frac{\sqrt {-20\,{{p_4}}^{2}+10\,{p_4}+10}}{3})g_2\\
8 \pi P_2 &=& (p_4+(\frac{1}{2}-\frac{\sqrt{5}}{50})\sqrt {-20\,{{p_4}}^{2}+10\,{p_4}+10})g_1 - \nonumber \\ & & (\frac{p_4}{3}+\frac{2}{3}+\frac{\sqrt {-20\,{{p_4}}^{2}+10\,{p_4}+10}}{3})g_2\\
8 \pi P_3 &=& (p_4+(\frac{1}{2}+\frac{\sqrt{5}}{50})\sqrt {-20\,{{p_4}}^{2}+10\,{p_4}+10})g_1 - \nonumber \\ & & (\frac{p_4}{3}+\frac{2}{3}+\frac{\sqrt {-20\,{{p_4}}^{2}+10\,{p_4}+10}}{3})g_2
\end{eqnarray}

where:

\begin{eqnarray}
g_1 &=& f_1 f_2 f_3 \left( \cosh \left( 1/3\,\Lambda\, \left( {p_4}+2+\sqrt {-4\,{{p_4}}^{2
}+2\,{p_4}+2}\sqrt {5} \right) t \right) \right) \\
g_2 &=& f_1 f_2 f_3 \left( \sinh \left( 1/3\,\Lambda\, \left( {p_4}+2+\sqrt {-4\,{{p_4}}^{2
}+2\,{p_4}+2}\sqrt {5} \right) t \right) \right)
\end{eqnarray}

with:

\begin{eqnarray}
f_1 &=& {e^{3\,\Lambda\,t{p_4}+6\,\Lambda\,t \left( -1/3\,{p_4}+1/3+1/6
\,\sqrt {-20\,{{p_4}}^{2}+10\,{p_4}+10} \right) -2\,\Lambda\,t+{s_3}\,l-{s_3}\,l\sqrt {5}}}\\
f_2 &=& {\frac {\Lambda\,{p_4}}{2{{b_0}}^{2}{{c_0}}^{2} \left( 1/3\,{p_4}+2/3+1/3\,\sqrt {-20\,{{p_4}}^{2}+10\,{p_4}+10} \right) }
}\\
f_3 &=&   {\rm sech}^2 \left( \Lambda\, \left( 1/3\,{p_4}+2/3+1/3\,
\sqrt {-20\,{{p_4}}^{2}+10\,{p_4}+10} \right) t \right) 
\end{eqnarray}

Though the pressure is clearly oscillatory and anisotropic, we can define an effective function:
\begin{equation}
P_{eff}= \frac{1}{3} {\sum_{i=1}^3{P_i}}
\end{equation}

This yields the ``equation of state'':
\begin{equation}
\rho=3P_{eff}
\end{equation}

which is that of a hot photon gas, expanding anisotropically, as may have been the case in the very early universe.

\section{Conclusions}
\label{sec:conclusions}

We have examined the behavior of a 5D universe with spatial slices that possess Bianchi type-IX geometries and a non-compact fifth coordinate.  We have shown that this solution oscillates between generalized Kasner epochs, but not indefinitely, and therefore does not possess deterministic chaos.  Analyzing this solution within the context of induced matter theory, we have found an effective equation of state, resembling an anisotropically expanding photon gas.  In this manner, induced matter theory has provided a possible explanation for the lack of chaos in vacuum 5D Bianchi type-IX models:  as expressed in 4D form, the induced matter acts to suppress transitions.  

\section{Acknowledgements}

We wish to thank Paul Wesson for useful discussions.

\end{document}